\documentclass[aps,prd,twocolumn,floatfix,showpacs]{revtex4} 

\usepackage{graphics}
\usepackage{graphicx}

\usepackage{latexsym}
\usepackage[cp866]{inputenc}
\usepackage[english]{babel}  

\input epsf

\begin{document}

\title{\boldmath Line shape of $\psi(3770)$ in $e^+e^-\to D\bar D$}
\author{N.N. Achasov\footnote{achasov@math.nsc.ru}
and G.N. Shestakov\footnote{shestako@math.nsc.ru}
}\affiliation{Laboratory of  Theoretical Physics, S.L. Sobolev
Institute for Mathematics,\\ 630090, Novosibirsk, Russia}


\begin{abstract} Interference phenomena observed in the $\psi(3770)$
resonance region in the reactions $e^+e^-$\,$\to$\,$ D\bar D$ are
analyzed. To avoid ambiguities in the determination of the
$\psi(3770)$ resonance parameters, when analyzing data between the
$D\bar D$ and $D\bar D\pi$ thresholds, the amplitudes satisfying the
elastic unitarity requirement should be used. In the lack of
information on the $P$ wave of $D\bar D$ elastic scattering, the
$\psi(3770)$ parameters, determined by fitting the $e^+e^-$\,$\to
$\,$D\bar D$ data, can essentially depend on the model used for the
total contribution of the resonance and background. The selection of
the models can be toughened by comparing their predictions with the
relevant data on the shape of the $\psi(3770)$ peak in the
non-$D\bar D$ channels $e^+e^-$\,$\to $\,$\gamma\chi_{c0}$,
$J/\psi\eta$, $\phi\eta$, etc.
\end{abstract}


\maketitle

\section{INTRODUCTION}

The resonance $\psi(3770)$ was investigated in the reactions
$e^+e^-$\,$\to $\,$D\bar D$ by the MARK-I \cite{Ra,Pe}, DELCO
\cite{Ba}, MARK-II \cite{Sch}, BES \cite{Ab1,Ab2,Ab3,Ab3a,Ab4,Ab5,
Ab6,Ab7,Ab8}, CLEO \cite{Be1,Do,Be2}, {\it BABAR} \cite{Au1,Au2},
Belle \cite{Pa}, and KEDR \cite{To1,An1,To2,An2} Collaborations.
With increasing accuracy of measurements, there appeared indications
on an unusual shape of the $\psi(3770)$ peak, i.e., on possible
interference phenomena in its region
\cite{Ab3a,Ab4,Ab6,Ab7,Ab8,Au1,Au2,Pa,To1,An1,To2,An2,Ya,LQY,ZZ}.
Recently, the KEDR Collaboration noted \cite{An1,To2,An2} that the
parameters of the $\psi(3770)$ resonance become distinctly different
from those quoted by the Particle Data Group in the preceding
reviews (see, for example, Ref. \cite{PDG10}) if the data analysis
takes into account the interference between the $\psi(3770)$
production amplitude and the nonresonant $D\bar D$ production one.
In Refs. \cite{To2,An2}, two very different solutions for the
interfering resonance and background amplitudes were obtained
\cite{PDG12}. These solutions lead to the same energy dependence of
the cross section and are indistinguishable by the $\chi^2$
criterion. Ambiguities of this type in the interfering resonances
parameters determination were found in Ref. \cite{Bu}.

CLEO-c has now accumulated about 818\,pb$^{-1}$ \cite{Me} and BES
III about 2.9\,fb$^{-1}$ \cite{Li} integrated luminosity on the
$\psi(3770)$ peak for open charm physics investigations. Therefore,
from CLEO-c and BES III, one can also expect new data with very high
statistics on the shape of the $\psi(3770)$ resonance. In this
regard, we believe it is timely to discuss some dangers which are
hidden in the commonly used schemes for the description of the
$\psi(3770)$ peak.

Section II shows that the most precise current data on the
$e^+e^-$\,$\to$\,$D\bar D$ reaction cross section in the
$\psi(3770)$ region are hard to describe with a single $\psi(3770)$
resonance. In Sec. III, simple models for the isoscalar part of the
$D$ meson electromagnetic form factor, $F^0_D$, which determines the
amplitude $e^+e^-$\,$ \to$\,$D\bar D$ in the $\psi(3770)$ resonance
region, are constructed. The models take into account interference
between the resonance and background contributions and yield good
descriptions of the data. The form factor $F^0_D$ is constructed in
such a way as to guarantee at least at the model level the elastic
unitarity requirement. Information on the $P$ wave of $D\bar D$
elastic scattering could be a great help in constructing the $D$
meson electromagnetic form factor. However, such information is not
available. Therefore, it is reasonable that the $\psi(3770)$
resonance parameters, derived from fitting the $e^+e^-$\,$\to$\,$ D
\bar D$ data, can essentially depend on the model used for the sum
contribution of the resonance and background. Section IV shows that
the selection of the models can be significantly toughened by
comparing their predictions with the relevant data on the shape of
the $\psi(3770)$ peak in the non-$D\bar D$ channels, such as
$e^+e^-$\,$\to$\,$\gamma\chi_{c0}$, $J/\psi\eta$, $\phi\eta $, etc.
The results of our analysis are briefly formulated in Sec. V. A
comment concerning the ambiguity of the fitting solutions found in
Refs. \cite{To2,An2,Bu} is given in the Appendix.

\section{\boldmath THE $\psi(3770)$ RESONANCE IN $e^+e^-\to D\bar D$}

\begin{figure}\hspace{-7mm}\centerline{\epsfysize=3.3in
\epsfbox{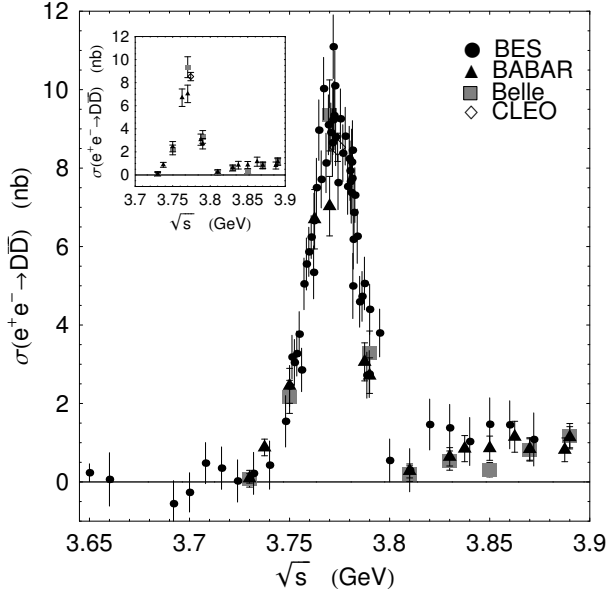}} 
\caption{The data from BES \cite{Ab3a,Ab4}, CLEO \cite{Be2}, {\it
BABAR} \cite{Au1,Au2}, and Belle \cite{Pa} for
$\sigma(e^+e^-$\,$\to$\,$ D\bar D)$. For clarity, the inset shows
only the points from CLEO \cite{Be2}, {\it BABAR} \cite{Au1,Au2},
and Belle \cite{Pa}.} \label{Figure1}
\end{figure}

Figure \ref{Figure1} shows the data for the sum of the
$e^+e^-$\,$\to$\,$D^0\bar D^0$ and $e^+e^-$\,$\to$\,$D^+D^-$
reaction cross sections in the $\psi(3770)$ region,
$\sigma(e^+e^-$\,$\to$\,$D\bar D)$, obtained by BES \cite{Ab3a,Ab4}
(68 points in the $\sqrt{s}$ region from 3.645 to 3.872\,GeV), CLEO
\cite{Be2} (1 point at $\sqrt{s}$\,=\,3.774\,GeV), {\it BABAR}
\cite{Au1,Au2} (15 points for 3.73\,GeV\,$<$\,$\sqrt{s}$\,$<
$\,3.89\,GeV), and Belle \cite{Pa} (9 points for 3.73\,GeV\,$<
$\,$\sqrt{s}$\,$<$\,3.89\,GeV). Here $\sqrt{s}$ is the energy in the
$D\bar D$ center-of-mass system. This is the most detailed and
accurate current data. It is clear, however, that further
improvement of the data and matching the results from different
groups are required \cite{FNNew}.

Note that the BES Collaboration \cite{Ab3a,Ab4} measured, in the
region up to the $D\bar D^*$ threshold ($\approx$\,3.872\,GeV), the
quantity $R(s)$\,=\,$\sigma(e^+e^-\to \mbox{hadrons})/\sigma(e^+e^-
\to\mu^+\mu^-)$ [where $\sigma (e^+e^-\to\mu^+\mu^-)$\,=\,$4\pi
\alpha^2/3s$ and $\alpha$\,=\,$1/137$]. The $D\bar D$ events were
not specially identified. The BES points shown in Fig. \ref{Figure1}
correspond to the cross section $(4\pi\alpha^2/3s )[R(s)-R_{uds}]$,
where $R_{uds}$\,=\,2.121 \cite{Ab4} describes the background from
the light hadron production. This cross section gives a good
estimate for $\sigma(e^+e^-$\,$\to$\,$D\bar D)$ in the $\psi(3770)$
region, because it is expected that the decay width of the $c\bar c$
state $\psi(3770)$ (or states) into the non-$D\bar D$ modes must be
comparable with the decay width of the $\psi(2S)$ resonance located
under the open charm production threshold. Hence, the ratio
$B(\psi(3770)\to\mbox{non-}D\bar D)/B(\psi(3770)\to D\bar D)$ must
be small owing to the large total width of the $\psi(3770)$. This is
confirmed by experiment \cite{PDG12,FN1}.

The measured $D\bar D$ mass spectrum \cite{Ab4,Au1,Au2,Pa} has the
following features. First, the right side of the $\psi(3770)$ peak
turns out to be more steep than its left side. Second, there is a
deep dip near 3.81\,GeV in the mass distribution (in fact, the cross
section dips to zero near this point). These features are hard to
describe with the help of a single $\psi(3770)$ resonance
contribution.

In most experimental works, the $e^+e^-\to D\bar D$ cross section
caused by the $\psi(3770)$ resonance production was described with
minor modifications by the following formula [below, for short
$\psi(3770)$ is also denoted as $\psi''$]:
\begin{equation}\label{Sigm1} \sigma_{\psi''}(e^+e^-\to D
\bar D)=\frac{12\pi\Gamma_{\psi''e^+e^-}\Gamma_{\psi''D\bar
D}(s)}{(m^2_{\psi''}-s)^2+(m_{\psi''}\Gamma^{tot}_{\psi''}(s))^2},
\end{equation} where $m_{\psi''}$ is the mass, $\Gamma_{\psi''e^+
e^-}$ the $e^+e^-$ decay width, $\Gamma_{\psi''D\bar D}(s)$ the
$D\bar D$ decay width, and $\Gamma^{tot}_{\psi''}(s)$ the total
decay width of the resonance. The energy-dependent width
$\Gamma_{\psi''D\bar D}(s)$ was taken in the form
\begin{equation}\label{GDD1} \Gamma_{\psi''D\bar D}(s)=G^2_{\psi''}
\left(\frac{p^3_0(s)} {1+r^2p^2_0(s)}+\frac{p^3_+(s)}
{1+r^2p^2_+(s)}\right),
\end{equation} where $p_0(s)$\,=\,$\sqrt{s/4- m^2_{D^0}}$ and
$p_+(s)$\,=\,$ \sqrt{s/4-m^2_{D^+}}$ are the $D^0$ and $D^+$
momenta, respectively, $r$ is the $D \bar D$ interaction radius
\cite{BW}, and $G_{\psi''}$ is the coupling constant of the $\psi''$
to $D\bar D$. Because the $\psi''$\,$\to$\,$D\bar D$ decay is
dominant \cite{PDG12}, we put in Eq. (\ref{Sigm1})
$\Gamma^{tot}_{\psi''}(s) $\,=\,$\Gamma_{\psi'' D\bar D}(s)$. This
simplification is not essential for our analysis.

\begin{figure}\hspace{-7mm}\centerline{\epsfysize=3.3in
\epsfbox{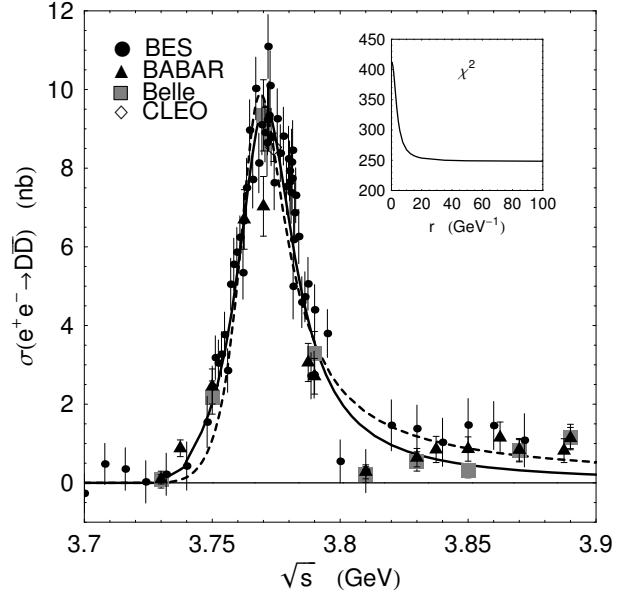}} 
\caption{The results of the fit using Eqs. (\ref{Sigm1}) and
(\ref{GDD1}). The illustration of the $\psi(3770)$ resonance shape
dependence on the parameter $r$. See the text for details.}
\label{Figure2}
\end{figure}

The dashed and solid curves in Fig. \ref{Figure2} show the fits to
the data in the region 3.72\,GeV\,$<$\,$\sqrt{s}$\,$<$\,3.9\,GeV (87
points) with the use of Eqs. (\ref{Sigm1}) and (\ref{GDD1}) at
$r$\,=\,0 and 100 GeV$^{-1}$, respectively. In the inset in this
figure, the quantity $\chi^2$, characterizing the goodness of fit,
is shown as a function of $r$. As $r$ increases from 0 approximately
to 15\,GeV$^{-1}$ ($\approx$\,3\,fm), the $\chi^2$ value sharply
decreases and then, with $r$ increasing, remains practically
unchanged. Such a behavior of $\chi^2$ leaves the parameter $r$ very
uncertain. Of course, too large values of $r$ hardly have any
physical means \cite{FN1a}. Owing to the parameter $r$ in
$\Gamma_{\psi''D\bar D}(s)$, one succeeds in raising the left side
of the $\psi(3770)$ peak and lowering its right side. In fact, all
existing data require such a deformation of the $\psi''$ peak.
However, as is seen from Fig. \ref{Figure2}, a dip near 3.81\,GeV
cannot be explained by varying $r$. The obtained very unsatisfactory
$\chi^2$ values (for the dashed and solid curves in Fig.
\ref{Figure2}, $\chi^2/n\mbox{\,d.o.f.}\approx413/84\approx4.9$ and
$248/83\approx3$, respectively) are due to both notable differences
between the data from different groups and the existence of the dip
in the $D\bar D$ mass spectrum (for example, for the solid curve in
Fig. \ref{Figure2}, the points at $\sqrt{s}$\,=\,3.8 and 3.81 GeV
yield $\chi^2\approx81$). In order to qualitatively improve the data
description in the $\psi''$ resonance region, in particular, to
explain a dip near 3.81\,GeV, it is necessary to take into account
the interference between the resonant and nonresonant $D\bar D$
production.

\section{\boldmath THE $D$ MESON ELECTROMAGNETIC FORM FACTOR}
\subsection{Unitarity requirement}

In constructing the model describing the process $e^+e^-$\,$\to$\,$D
\bar D$ one must keep in mind that we investigate the $D$ meson
electromagnetic form factor, the phase of which in the elastic
region is completely fixed by the unitarity condition (or the Watson
theorem of final-state interaction). Experiment clearly indicates
that we deal with the resonant scattering of $D$ mesons. Really,
there is the $\psi''$ resonance between the $D\bar D$ and $D\bar
D^*$ thresholds ($2m_D\approx3.739$\,GeV and $m_D+m_{D^*}\approx3.
872$\,GeV), which in a good approximation can be considered as an
elastic one, because it has no appreciable non-$D\bar D$ decays
\cite{PDG12}. Usually, such scattering is described as resonance
scattering with an elastic background --- see, for example, Ref.
\cite{ADS84} --- i.e., the corresponding strong amplitude $T^I_J$
with the definite isospin $I$ and spin $J$ (in our case, it is the
$D\bar D$ scattering amplitude $T^0_1$) is given by \cite{FN2}
\begin{equation}\label{T10} T^0_1=e^{i\delta^0_1}\sin\delta^0_1=
\frac{e^{2i\delta_{bg}}-1}{2i}+e^{2i\delta_{bg}}T_{res}\,,
\end{equation} where $\delta^0_1=\delta_{bg}+\delta_{res}$
is the scattering phase, $\delta_{bg}$ is the elastic background
phase (or the phase of potential scattering), and $\delta_{res}$ is
the phase of the resonance amplitude $T_{res}$ (in the simplest
parametrization $T_{res}$\,=\,$\frac{\Gamma/2}{M-E-i \Gamma/2}$).
Then, according to the unitarity condition $\mbox{Im}F^0_D$\,=\,$
F^0_D \,T^{0*}_1$, the $D$ meson isoscalar form factor $F^0_D$
\cite{FN3} has the form in the elastic region
\begin{equation}\label{S2FD0} F^0_D=e^{i\delta^0_1}G^0_D
=e^{i(\delta_{bg}+\delta_{res})}G^0_D\,,\end{equation} where $G^0_D$
is the real function of energy. A similar representation of the
amplitude $e^+e^-$\,$\to$\,$D\bar D$ used for the data description
guarantees the unitarity requirement on the model level. The sum of
the $e^+e^-$\,$\to$\,$D\bar D$ reaction cross sections is expressed
in terms of $F^0_D$ in the following way:
\begin{equation}\label{Sigm2} \sigma^{D\bar
D}(s)= \frac{8\pi\alpha^2}{3s^{5/2}}\left|F^0_D(s)\right|^2
\left[p^3_0(s)+p^3_+(s)\right].\end{equation}

\begin{figure}\hspace*{2.4mm}\centerline{\epsfysize=2.4in
\epsfbox{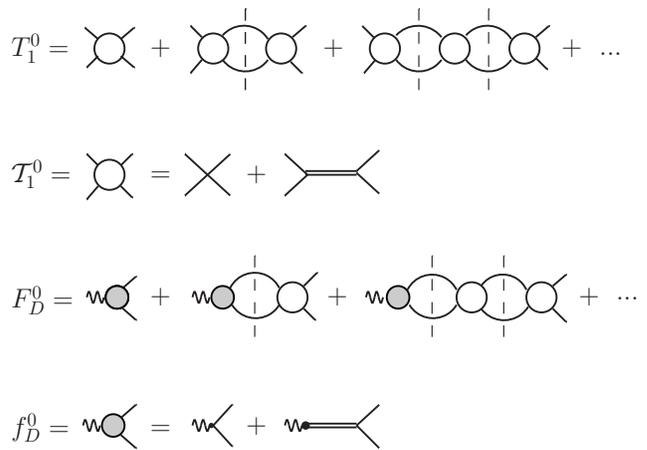}} 
\caption{The graphical representation of the strong $D\bar D$
scattering amplitude $T^0_1$ and the $D$ meson electromagnetic form
factor $F^0_D$. The vertical dashed lines show that the $D$ and
$\bar D$ mesons in the loops are on the mass shell. Diagrams
corresponding to the amplitude $\mathcal{T}^0_1$ and the form factor
$f^0_D$ show the structure of primary mechanisms included in the
model to describe the $\psi''$ resonance region.}
\label{Figure3a}\end{figure}

\subsection{\boldmath A simplest model for $F^0_D$: Resonance plus background}

To understand how the form factor and strong amplitude can be
constructed to satisfy the unitarity requirement, the easiest way to
use the field-theory model shown in Fig. \ref{Figure3a} and write
\begin{equation}\label{T10mod1}
 T^0_1(s)=\frac{\mathcal{T}^0_1(s)}{1-i\mathcal{T}^0_1(s)}\,,
\end{equation}
\begin{equation}\label{F0Dmod1}
 F^0_D(s)=\frac{f^0_D(s)}{1-i\mathcal{T}^0_1(s)}\,,
\end{equation} where
\begin{equation}\label{Ti10mod1}
 \mathcal{T}^0_1(s)=
 \nu(s)t^0_1(s)\,,
\end{equation}
\begin{equation}\label{nu1}
 \nu(s)=[p^3_0(s)+p^3_+(s)]/\sqrt{s}\,,
\end{equation}
\begin{equation}\label{t10mod1}
 t^0_1(s)=\lambda+\frac{1}{6\pi}\frac{g^2_{\psi''D\bar D}}
 {m^2_{\psi''}-s}\,,
\end{equation}
\begin{equation}\label{f0Dmod1}
 f^0_D(s)=\lambda_\gamma+\frac{g_{\psi''\gamma}g_{\psi''D\bar
 D}}{m^2_{\psi''}-s}\,.
\end{equation}
Graphically, the amplitude $T^0_1(s)$ and the form factor $F^0_D(s)$
defined in Eqs. (\ref{T10mod1}) and (\ref{F0Dmod1}) corresponds to
the infinite chains of the diagrams in Fig. \ref{Figure3a} with the
real $D$ and $\bar D$ mesons in the intermediate states. The
amplitude $t^0_1 $ and the form factor $f^0_D$ defined in Eqs.
(\ref{t10mod1}) and (\ref{f0Dmod1}) specify the structure of primary
mechanisms included in the model to describe the $\psi''$ resonance
region. The constants $\lambda$ and $\lambda_\gamma$ effectively
take into account background (nonresonant in the $\psi''$ region)
contributions to the strong amplitude and form factor, respectively,
and the constants $g_{\psi''D\bar D}$ and $g_{\psi''\gamma}$
describe couplings of the $\psi''$ to the $D\bar D$ and virtual
$\gamma$ quantum, respectively. The requirement of the unitarity
condition is fulfilled in the model under consideration: The phase
of the form factor $F^0_D(s)$ is defined by the phase of the
amplitude $T^0_1(s)$. This phase has the dynamical origin.

The physical content of Eqs. (\ref{T10mod1}) and (\ref{F0Dmod1})
will become more clear if they are rewritten in the form of Eqs.
(\ref{T10}) and (\ref{S2FD0}), respectively. As a result, we obtain
the following expressions for the background and resonance
components of $T^0_1(s)$:
\begin{equation}\label{Tbg} T_{bg}=
\frac{e^{2i\delta_{bg}(s)}-1}{2i}=\frac{\nu(s)\lambda}{1-i\nu(s)\lambda}\,,
\end{equation}
\begin{equation}\label{Tres} T_{res}=
\frac{\sqrt{s}\Gamma_{\psi''D\bar
D}(s)}{M^2_{\psi''}-s+\mbox{Re}\Pi_{\psi''}(M^2_{\psi''})-\Pi_{\psi''}(s)}\,,
\end{equation} where
\begin{equation}\label{ImPi} \mbox{Im}\Pi_{\psi''}(s)=\sqrt{s}\Gamma_{\psi''D\bar
D}(s)=\frac{\widetilde{g}^2_{\psi''D\bar D}(s)}{6\pi}\,\nu(s)\,,
\end{equation}
\begin{equation}\label{RePi} \mbox{Re}\Pi_{\psi''}(s)=-\lambda\frac{
\widetilde{g}^2_{\psi''D\bar D}(s)}{6\pi}\,\nu(s)^2\,,
\end{equation}
\begin{equation}\label{M2psi2} M^2_{\psi''}=m^2_{\psi''}-\mbox{Re}
\Pi_{\psi''}(M^2_{\psi''})\,,
\end{equation}
\begin{equation}\label{gpsi2DDs} \widetilde{g}_{\psi''D\bar D}(s)=\frac{
g_{\psi''D\bar D}}{|1-i\nu(s)\lambda|}\,.
\end{equation}
For $F^0_D(s)$ we obtain
\begin{equation}\label{F0Dmod1a}
F^0_D(s)=e^{i\delta^0_1(s)}\frac{(m^2_{\psi''}-s)\widetilde{\lambda}_\gamma(s)+g_{
\psi''\gamma}\widetilde{g}_{\psi''D\bar
D}(s)}{|M^2_{\psi''}-s+\mbox{Re}
\Pi_{\psi''}(M^2_{\psi''})-\Pi_{\psi''}(s)|}\,,
\end{equation} where $\delta^0_1(s)$\,=\,$\delta_{bg}(s)+\delta_{res}(s)$
[$\delta_{bg}(s)$ and $\delta_{res}(s)$ are the phases of the
amplitudes (\ref{Tbg}) and  (\ref{Tres}), respectively] and
\begin{equation}\label{lambdagammas} \widetilde{\lambda}_\gamma(s)=\frac{
\lambda_\gamma}{|1-i\nu(s)\lambda|}\,.
\end{equation}

\begin{figure}\hspace{-7mm}\centerline{\epsfysize=3.3in
\epsfbox{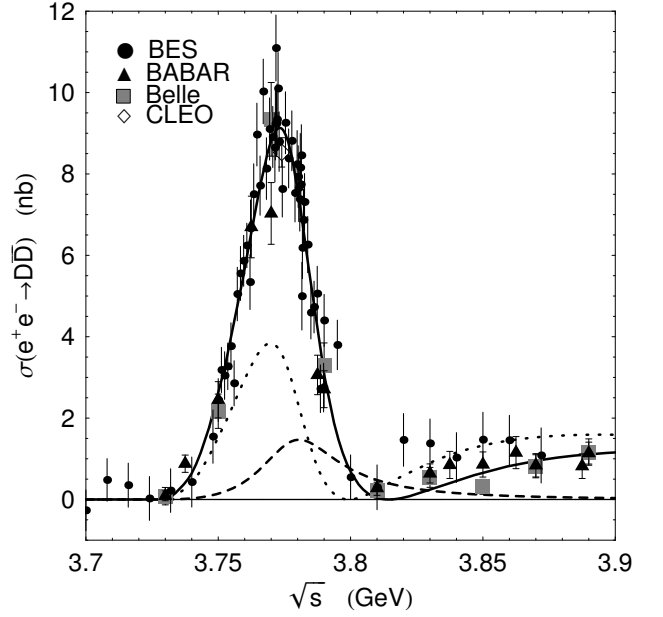}} 
\caption{The resonance plus background model. The solid curve is the
result of fitting the data with the use of Eqs. (\ref{Sigm2}) and
(\ref{F0Dmod1a}). The dashed curve shows the contribution from the
$\psi''$ resonance production ($\sim g_{\psi''\gamma}$ in $F^0_D$),
and the dotted curve shows the contribution from the background
production ($\sim \lambda_\gamma$ in $F^0_D$) modified by the strong
resonance and background final-state interactions.} \label{Figure4}
\end{figure}

Thus $F^0_D(s)$ incorporates the resonance contribution
(proportional to $g_{ \psi''\gamma}$) modified (dressed) by the
strong background \cite{A0} and the proper background contribution
(proportional to $\lambda_{\gamma}$) modified by the strong
resonance and background final-state interactions. The numerator in
Eq. (\ref{F0Dmod1a}) is proportional to the first-degree polynomial
in $s$, $\lambda_\gamma(m^2_{ \psi''}-s)+g_{\psi''\gamma}g_{\psi''
D\bar D}$, with real coefficients. This ensured that the dip in
$\sigma(e^+e^-$\,$\to $\,$D\bar D)$ near 3.81 GeV can be explained
by the zero in $F^0_D(s)$, caused by compensation between the
$\psi''$ resonance and background contributions. Note that the
presence of the zero in $F^0_D(s)$ is in qualitative agreement with
the coupled-channel model prediction \cite{EGKLY}.

As is seen from Fig. \ref{Figure4}, the constructed model for
$F^0_D(s)$ yields a quite reasonable description of the data (here
$\chi^2/n\mbox{\,d.o.f.}\approx123/82\approx1.5$, which is much
better than the above $\chi^2/n\mbox{\,d.o.f.}$ values for the fits
shown in Fig. \ref{Figure2}). For the solid curve in Fig.
\ref{Figure4}, the cross section at the maximum (located at
$\sqrt{s}$\,=\,$\sqrt{s_{max}}\approx3.773$ GeV)
$\sigma_{max}\approx9.13$ nb, the full width of the peak at its half
maximum $\Gamma_{hmax}\approx29.7$ MeV, and the effective electron
width of the resonance structure $\Gamma^{eff }_{e^+ e^-}$\,=\,$
s_{max}\sigma_{max}\Gamma_{hmax}/(12\pi) \approx0.263$ keV. These
characteristics of the observed peak are in close agreement with the
values of the mass ($\approx3.773$ GeV), the total width
($\approx27.2$ MeV), and the electron partial width ($\approx0.262$
keV) which are quoted by the Particle Data Group \cite{PDG12} as the
averaged individual characteristics of the $\psi''$ resonance.
However, the peak (in its line shape there is a zero at
$\sqrt{s}\approx 3.814 $\,GeV) does not correspond to a solitary
resonance. Therefore, it is reasonable that the model parameters for
$\psi''$ differ from the effective parameters of the visible peak.
Let us present the corresponding numbers.

The curves in Fig. \ref{Figure4} correspond to the following values
of the fitted parameters: $m_{\psi''}$\,=\,3.799 GeV,
$g_{\psi''D\bar D}$\,=\,$\pm$\,19.35,
$g_{\psi''\gamma}$\,=\,$\pm$\,0.1483 GeV$^2$, $\lambda
$\,=\,$-30.35$ GeV$^{-2}$, and $\lambda_\gamma$\,=\,$\pm$\,25.07 [if
$\lambda_\gamma$\,$>$\,0 ($<$\,0), then
$g_{\psi''\gamma}g_{\psi''D\bar D}$\,$>$\,0 ($<$\,0); see Eq.
(\ref{F0Dmod1a})]. As the individual characteristics of the $\psi''$
resonance, one can take the quantities dressed (renormalized) by the
background contributions [see Eqs. (\ref{Tres})--(\ref{gpsi2DDs})]:
$M_{\psi''}$\,=\,3.784 GeV, $\Gamma^{ren}_{\psi''D\bar
D}$\,=\,$\Gamma_{\psi''D\bar D}(M^2_{\psi''})/Z_{\psi''}$\,=\,37.61
MeV, and $\Gamma^{ren}_{\psi''e^+e^-}$\,=\,$\Gamma_{\psi''e^+
e^-}/Z_{\psi''}$\,=\,0.05181 keV, where $Z_{\psi''}$\,=\,$1+
\mbox{Re}\Pi'_{\psi''}(M^2_{\psi''})$\,=\,1.748 and $\Gamma_{
\psi''e^+ e^-}$ $=4\pi\alpha^2g^2_{\psi''\gamma}/(3M^3_{\psi''})$.

The obvious drawback of the considered model is the uncertain nature
of the background contributions. Therefore, the validity of this
model is hard to verify in other reactions. However, the model can
be easily improved. It is clear that the main sources of the
background in the $\psi''$ region are the tails from the $J/\psi$,
$\psi(2S)$, $\psi(4040)$, $\psi(4160)$, and other resonances. The
right number of resonances can be incorporated in the model by
adding the corresponding pole terms to expressions (\ref{t10mod1})
and (\ref{f0Dmod1}) for $t^0_1(s)$ and $f^0_D(s)$. In that case, the
parameters $\lambda$ and $\lambda_\gamma$ will effectively describe
the contributions from the residual background, and it is hoped that
they will be small. The $\psi(2S)$ resonance is closest to the
$\psi''$. Its coupling to $e^+e^-$ is about an order of magnitude
larger than that of $\psi''$ \cite{PDG12}, and there are no apparent
reasons for the suppression of the coupling of the $\psi(2S)$ to
$D\bar D$. In the next subsection, we will consider in detail the
model taking into account the $\psi(2S)$ resonance contribution and,
in Sec. IV, discuss additional ways of checking this model.

\subsection{\boldmath The model for $F^0_D$ with the $\psi''$ and
$\psi(2S)$ resonances}

The connection of the $\psi(2S)$ contribution does not change the
structure of expressions (\ref{T10mod1}) and (\ref{F0Dmod1}) for
$T^0_1$ and $F^0_D$. Only the functions $t^0_1$ and $f^0_D$ change.
Now they are given by
\begin{equation}\label{t10mod2}
 t^0_1(s)=\lambda+\frac{1}{6\pi}\frac{g^2_{\psi(2S)D\bar D}}{m^2_{\psi(2S)}-s}
 +\frac{1}{6\pi}\frac{g^2_{\psi''D\bar D}}{m^2_{\psi''}-s}\,,
\end{equation}
\begin{equation}\label{f0Dmod2}
 f^0_D(s)=\lambda_\gamma+\frac{g_{\psi(2S)\gamma}g_{\psi(2S)D\bar
 D}}{m^2_{\psi(2S)}-s}+\frac{g_{\psi''\gamma}g_{\psi''D\bar
 D}}{m^2_{\psi''}-s}\,.
\end{equation}
Hereinafter we use the values of $m_{\psi(2S)}=3.6861$\,GeV
\cite{PDG12} and $\Gamma_{\psi(2S)e^+e^-}=2.35$\,keV \cite{PDG12}.
From the relation $\Gamma_{\psi(2S)e^+e^-}=4\pi \alpha^2g^2_{\psi
(2S)\gamma}/(3m^3_{\psi(2S)})$, we get $g_{\psi(2S)\gamma}
\approx\pm0.7262$ GeV$^2$. The coupling constant $g_{\psi(2S)D\bar
D}$ is a free parameter.

Owing to the common $D^0\bar D^0$ and $D^+D^-$ decay channels, the
$\psi''$ and $\psi(2S)$ resonances can transform into each other
(i.e., mix); for example, $\psi''$\,$\to$\,$D\bar D$\,$\to$\,$\psi
(2S)$. Therefore, it is very useful to rewrite Eqs. (\ref{T10mod1})
and (\ref{F0Dmod1}) for the amplitude $T^0_1$ and the form factor
$F^0_D$ in terms which would reflect this physical aspect of the
model and, in particular, introduce the amplitude describing the
$\psi''-\psi(2S)$ mixing.

Let us write the background amplitude in the form similar to Eq.
(\ref{Tbg}):
\begin{equation}\label{Tbg1} T_{bg}=
\frac{e^{2i\delta_{bg}(s)}-1}{2i}=\frac{\nu(s)\lambda}{1-i\nu(s)\lambda}\,.
\end{equation}
The amplitude $T_{res}$ [see Eq. (\ref{T10}], corresponding to the
complex of the mixed $\psi''$ and $\psi(2S)$ resonances, dressed by
the residual background, we represent in the following symmetric
form \cite{A1,A2,A3}:
\begin{equation}\label{FresMix}
T_{res}=\frac{(m^2_{\psi''}-s)\mbox{Im}\Pi_{\psi(2S)}
(s)+(m^2_{\psi(2S)}-s)\mbox{Im}\Pi_{\psi''}(s)}{D_{\psi''}(s)D_{\psi
(2S)}(s)-\Pi^2_{\psi''\psi(2S)}(s)}\,,
\end{equation} where $D_{\psi''}(s)$ and $D_{\psi(2S)}(s)$ are the inverse
propagators of $\psi''$ and $\psi(2S)$, respectively,
\begin{equation}\label{Dpsi2}
D_{\psi''}(s)=m^2_{\psi''}-s-\Pi_{\psi''}(s)\,,
\end{equation}
\begin{equation}\label{Dpsi1}
D_{\psi(2S)}(s)=m^2_{\psi(2S)}-s-\Pi_{\psi(2S)}(s)\,,
\end{equation}
\begin{equation}\label{POpsi2}
\Pi_{\psi''}(s)=\frac{i}{6\pi}\frac{g^2_{\psi''D\bar
D}}{1-i\nu(s)\lambda}\,\nu(s)\,,
\end{equation}
\begin{equation}\label{POpsi1}
\Pi_{\psi(2S)}(s)=\frac{i}{6\pi}\frac{g^2_{\psi(2S)D\bar
D}}{1-i\nu(s)\lambda}\,\nu(s)\,,
\end{equation} and $\Pi_{\psi''\psi(2S)}(s)$ is the amplitude describing the
$\psi''-\psi(2S)$ mixing caused by the $\psi''$\,$\to$\,$ D\bar
D$\,$\to$\,$\psi(2S)$ transitions via the real $D\bar D$
intermediate states,
\begin{equation}\label{POpsi2psi1}
\Pi_{\psi''\psi(2S)}(s)=\frac{i}{6\pi}\frac{g_{\psi''D\bar
D}g_{\psi(2S)D\bar D}}{1-i\nu(s)\lambda}\,\nu(s)\,.
\end{equation}
Note that the phase of $T_{res}$ is defined by that of the
denominator in Eq. (\ref{FresMix}).

For the form factor, we get
\begin{equation}\label{F0DMix}
F^0_D(s)=e^{i\delta_{bg}(s)}\,\frac{\mathcal{R}_{D\bar
D}(s)}{D_{\psi''}(s)D_{\psi (2S)}(s)-\Pi^2_{\psi''\psi(2S)}(s)}\,,
\end{equation} where
\begin{eqnarray}\label{RDDMix1}
& \mathcal{R}_{D\bar D}(s)=(m^2_{\psi''}-s)(m^2_{\psi(2S)}-s)
\widetilde{\lambda}_\gamma(s) & \nonumber\\ & +g_{\psi(2S)
\gamma}[D_{\psi''}(s)\widetilde{g}_{\psi(2S)D\bar D}(s)+
\Pi_{\psi''\psi(2S)}(s)\widetilde{g}_{\psi''D\bar D}(s)] &  \nonumber\\
& +g_{\psi''\gamma}[D_{\psi(2S)}(s)\widetilde{g}_{\psi''D\bar D}(s)+
\Pi_{\psi''\psi(2S)}(s)\widetilde{g}_{\psi(2S)D\bar D}(s)] & \nonumber\\
&  & \end{eqnarray} and after cancellations
\begin{eqnarray}\label{RDDMix2}
& \mathcal{R}_{D\bar D}(s)=(m^2_{\psi''}-s)
(m^2_{\psi(2S)}-s)\widetilde{\lambda}_\gamma(s) & \nonumber\\
& +(m^2_{\psi''}-s)g_{\psi(2S)\gamma}\widetilde{g}_{\psi(2S)D\bar
D}(s) & \nonumber\\ & +(m^2_{\psi(2S)}-s)g_{\psi''\gamma}
\widetilde{g}_{\psi''D\bar D}(s). &
\end{eqnarray} Here $\widetilde{g}_{\psi(2S)D\bar D}(s)$\,=\,$ g_{\psi(2S)D\bar
D}/|1-i\nu(s)\lambda|$; $\widetilde{g}_{\psi''D\bar D}(s)$ and
$\widetilde{\lambda}_\gamma(s)$ are given by Eqs. (\ref{gpsi2DDs})
and (\ref{lambdagammas}), respectively.

\begin{figure}\hspace{-7mm}\centerline{\epsfysize=3.3in
\epsfbox{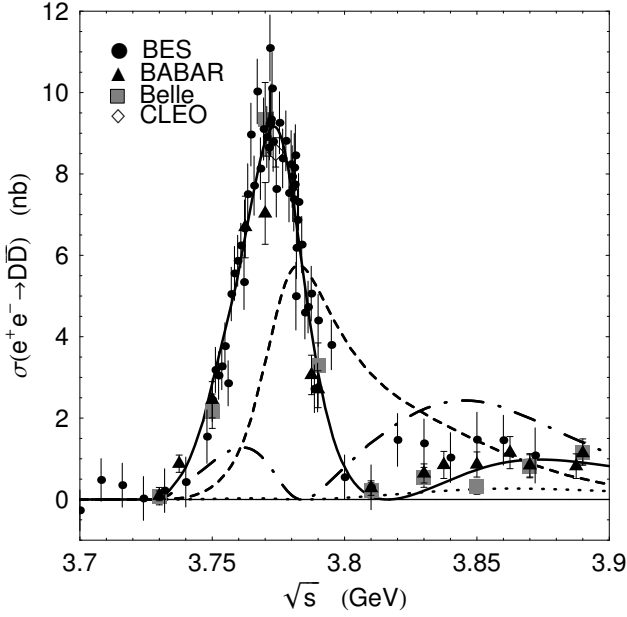}} 
\caption{The model with the $\psi''$ and $\psi(2S)$ resonances. The
solid curve is the fit using Eqs. (\ref{Sigm2}) and (\ref{F0DMix}).
The dashed, dot-dashed, and dotted curves show the $\psi''$,
$\psi(2S)$, and background production contributions proportional to
the coupling constants $g_{\psi''\gamma}$, $g_{\psi(2S)\gamma}$, and
$\lambda_\gamma$ in Eq. (\ref{RDDMix2}), respectively.}
\label{Figure5}\end{figure}

The curves in Fig. \ref{Figure5} correspond to the following values
of the fitted parameters: $m_{\psi''}$\,=\,3.784 GeV,
$g_{\psi''D\bar D}$\,=\,$\pm$\,13.21,
$g_{\psi''\gamma}$\,=\,$\pm$\,0.2237 GeV$^2$, $g_{\psi(2S)D\bar
D}$\,=\,$\pm$\,12.91, $\lambda $\,=\,$26.89$ GeV$^{-2}$, and
$\lambda_\gamma$\,=\,$\pm$\,2.456 [if $\lambda_\gamma$\,$>$\,0
($<$\,0), then $g_{\psi''\gamma}g_{\psi''D\bar D}$\,$>$\,0 ($<$\,0)
and $g_{\psi(2S)\gamma}g_{\psi(2S)D\bar D}$\,$<$\,0 ($>$\,0); see
Eq. (\ref{RDDMix2})]. Note that here $|\lambda_\gamma|$ is about an
order of magnitude smaller than in the previous case, as
qualitatively expected. For this fit, $\chi^2/n\mbox{\,d.o.f.}
\approx125/81\approx1.54$. The form factor has the zero at
$\sqrt{s}\approx 3.816$\,GeV.

Notice that the above estimates of $g_{\psi''D\bar D}$ and
$g_{\psi(2S)D \bar D}$ are in agrement with the corresponding values
obtained in the previous works utilizing other phenomenological
approaches \cite{Ya,LQY,ZZ}. For instance Ref. \cite{Ya}, from the
branching ratio of $\psi''$\,$\to$\,$D^0\bar D^0,D^+D^-$, gives
$g_{\psi''D\bar D}$\,=\,12.7, Ref. \cite{LQY} has $g_{\psi''D\bar
D}$\,=\,12.8 and $g_{\psi(2S)D\bar D}$\,=\,12, and Ref. \cite{ZZ}
fits $g_{\psi''D\bar D}$\,=\,$13.58\pm1.07$ and $g_{\psi(2S)D\bar
D}$\,=\,$9.05\pm2.34$ from $e^+e^-$\,$\to $\,$D^0\bar D^0$ and
$g_{\psi''D\bar D}$\,=\,$10.71\pm1.75$ and $g_{\psi(2S)D\bar
D}$\,=\,$7.72\pm1.02$ from $e^+e^-$\,$\to$\,$D^+ D^-$.

As the individual characteristics of the $\psi''$ resonance, one can
take again the quantities dressed (renormalized) by the background
contributions: $M_{\psi''}$\,=\,3.789 GeV,
$\Gamma^{ren}_{\psi''D\bar D}$\,=\,$\Gamma_{\psi''D\bar
D}(M^2_{\psi''})/Z_{\psi''}$\,=\,58.03 MeV, and
$\Gamma^{ren}_{\psi''e^+e^-}$\,=\,$\Gamma_{\psi''e^+ e^-}/Z_{
\psi''}$\,=\,0.2973 keV, where $Z_{\psi''}$\,=\,$1+\mbox{Re}\Pi'_{
\psi''}(M^2_{\psi''})$\,=\,0.6905 and $\Gamma_{\psi''e^+ e^-}=4\pi
\alpha^2g^2_{\psi''\gamma}/(3M^3_{\psi''})$. We calculated the above
parameters with the use of Eqs. (\ref{Tbg})--(\ref{gpsi2DDs}) by
making the substitution
\begin{equation}\label{Change}
\lambda\,\to\,\lambda+\frac{1}{6\pi}\frac{g^2_{\psi(2S)D\bar
D}}{m^2_{\psi(2S)}-s}\,,
\end{equation} i.e., in the $\psi''$ resonance region, we included
in $T_{bg}$ the total background from the amplitude $\lambda$ and
the $\psi(2S) $ contribution and took into account in $T_{res}$ the
$\psi''$ contribution dressed by this total background. For example,
at the $D\bar D$ threshold, $\lambda+\frac{1}{6\pi}\frac{g^2_{
\psi(2S)D\bar D}}{m^2_{\psi(2S)}-4m^2_D}\approx4.38$ GeV$^{-2}$
instead of $\lambda\approx-30.35$ GeV$^{-2}$ in the resonance plus
background model.

\begin{figure}\hspace{-7mm}\centerline{\epsfysize=5.2in
\epsfbox{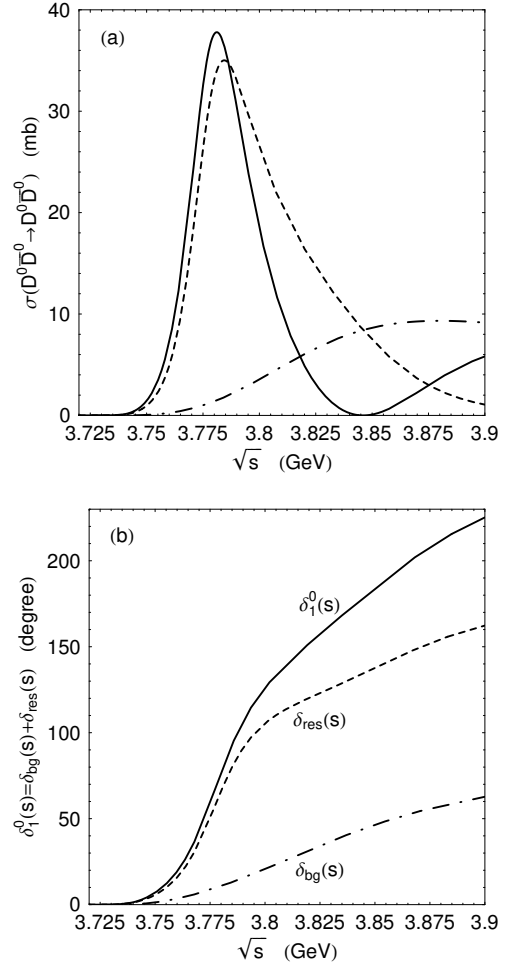}} 
\caption{The predictions of the model with the $\psi''$ and
$\psi(2S)$ resonances. (a) The cross sections and (b) the phase
shifts of $D\bar D$ elastic scattering in the $P$ wave. In (a), the
solid, dashed, and dot-dashed curves correspond to $\sigma(D^0\bar
D^0\to D^0\bar D^0)$\,=\,$3\pi|\sin\delta^0_{1} (s)|^2/p^2_0(s)$,\,
$3\pi|\sin\delta_{res}(s)|^2/p^2_0(s)$, and
$3\pi|\sin\delta_{bg}(s)|^2/p^2_0(s)$, respectively. In particular,
the model predicts that $\sigma(D^0\bar D^0$\,$\to$\,$D^0\bar
D^0)$\,=\,0 ($\delta^0_{1}$\,=\,$180^\circ$) at $\sqrt{s}\approx
3.846$ GeV.} \label{Figure5b}
\end{figure}

Thus, the fitting of the mass spectrum in $e^+e^-\to D\bar D$
permits us to determine the resonance and background characteristics
in specific models. Nevertheless, the information only on the
reactions $e^+e^-\to D\bar D$ is still lacking to give reliable
conclusions about the separate components of the reaction amplitude.
The performed analysis indicates that these components can be very
different in the different models. On the other hand, it is clear
that the interference pattern in the $\psi''$ region depends on the
reaction. Therefore, to toughen the selection of the models one
should compare their predictions with the experimental data on the
mass spectra for several different reactions.

For example, after the fitting of the $e^+e^-\to D\bar D$ data we
all know about $D\bar D$ elastic scattering in the $P$ wave at the
model level; see Fig. \ref{Figure5b}. Unfortunately, these
predictions are not possible to verify. However, there are many
other reactions which can be measured experimentally.


\section{\boldmath THE $\psi''$ SHAPE IN NON-$D\bar D$ DECAY CHANNELS}

Now we apply the last described model to construct the mass spectra
in the reactions $e^+e^-$\,$\to$\,$\gamma\chi_{c0} $, $J/\psi\eta$,
$\phi\eta$. In the $\psi''$ region, we restrict ourselves to the
contributions only from the $\psi''$ and $\psi(2S)$ resonances,
taking into account their couplings to the $\gamma\chi_{c0}$,
$J/\psi\eta$, and $\phi\eta$ channels in the first order of
perturbation theory.

The cross section for $e^+e^-$\,$\to$\,$ab$
($ab$\,=\,$\gamma\chi_{c0}$, $J/\psi\eta$, $\phi\eta$) can be
written as
\begin{equation}\label{SigmAB} \sigma^{ab}(s)=\frac{4\pi\alpha^2k^3_{ab}(s)}
{3s^{3/2}}\left|F_{ab}(s)\right|^2\,,\end{equation} where
$k_{ab}(s)$\,=\,$\sqrt{[s-(m_a+m_b)^2][s-(m_a-m_b)^2]}\,/(2\sqrt{s})$
and the form factor
\begin{equation}\label{FAB}
F_{ab}(s)=\frac{\mathcal{R}_{ab}(s)}{D_{\psi''}(s)D_{\psi
(2S)}(s)-\Pi^2_{\psi''\psi(2S)}(s)}\,,
\end{equation} where
\begin{eqnarray}\label{RAB} 
& \mathcal{R}_{ab}(s)=g_{\psi(2S)\gamma}[D_{\psi''}(s)g_{
\psi(2S)ab}+\Pi_{\psi''\psi(2S)}(s)g_{\psi''ab}] & \nonumber\\
& +g_{\psi''\gamma}[D_{\psi(2S)(s)}g_{\psi''ab}+
\Pi_{\psi''\psi(2S)}(s)g_{\psi(2S)ab}] &\end{eqnarray} and
$g_{\psi(2S)ab}$, $g_{\psi''ab}$ are the coupling constants of the
$\psi(2S)$, $\psi''$ to the $ab$ channel.

Table I presents information about the $\psi(2S)$ and $\psi''$
resonances in the $\gamma\chi_{c0}$, $J/\psi\eta$, and $\phi\eta$
decay channels \cite{PDG12,Br,Am,As}, which we use to construct the
corresponding mass spectra. The values for $g_{\psi(2S)ab}$
indicated in the table are obtained, up to the sign, from the data
on the $\psi(2S)$\,$\to$\,$ab$ decay widths by the formula
\begin{equation}\label{Gpsi2SAB}\Gamma_{\psi(2S)ab}=\frac{g^2_{\psi(2S)ab}}
{12\pi}k^3_{ab}(m^2_{\psi(2S)})\,.\end{equation} Note that the
available information about the $\psi''$\,$\to$\,$\gamma\chi_{c0}$,
$J/\psi\eta$, $\phi\eta$ decays are very poor \cite{PDG12}. Data on
the mass spectra in these channels are still absent.
\begin{table} 
\caption{Information about the $\psi(2S)$ and $\psi''$ resonances in
$\gamma\chi_{c0}$, $J/\psi\eta$, and $\phi\eta$ decay channels
\cite{PDG12,Br,Am,As} (errors $<$\,10\% are not shown).}
\vspace{0.05cm} 
\begin{tabular}{|c|c|c|c|}
\hline $ab$ channel & $\gamma\chi_{c0}$ & $J/\psi\eta$ & $\phi\eta$  \\
\hline $B(\psi(2S)\to ab)$ & 9.68\% & 3.28\% &
$(2.8^{+1.0}_{-0.8})\times10^{-5}$ \\
\hline $\Gamma_{\psi(2S)ab}$ (keV) & 29.4 & 10.0 & $(8.5^{+3.0}_{-2.4})\times10^{-3}$ \\
\hline $g_{\psi(2S)ab}$ (GeV$^{-1})$ & $\pm0.25$ & $\pm0.22$ & $\pm(2.7^{+0.5}_{-0.4})\times10^{-4}$ \\
\hline $\sigma(e^+e^-$\,$\to$\,$ab$) (pb); & $72\pm9$ & $8.6\pm3.9$ & $3.1\pm0.8$ \\
at $\sqrt{s}=3773$ MeV  &  &  &  \\
\hline\end{tabular}\end{table}
\begin{figure}\hspace{-7mm}\centerline{\epsfysize=2.6in 
\epsfbox{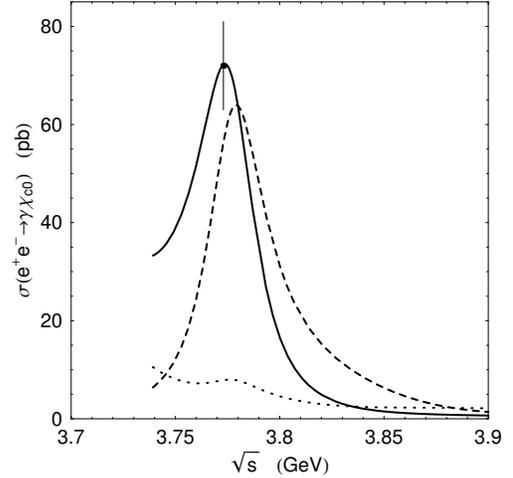}} 
\caption{The cross section for $e^+e^-$\,$\to$\,$\gamma\chi_{ c0}$.}
\label{Figure6}
\end{figure}
\begin{figure}\hspace{-7mm}\centerline{\epsfysize=2.6in
\epsfbox{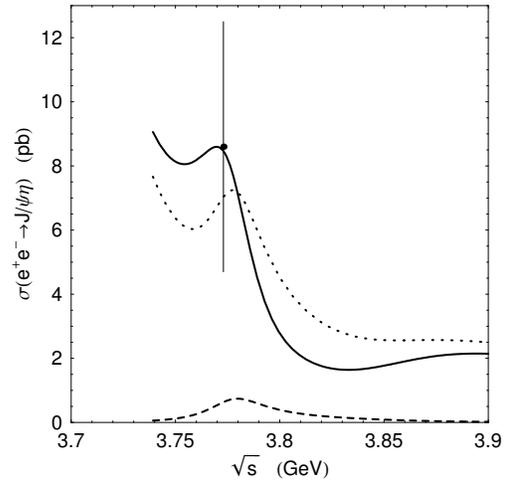}} 
\caption{The cross section for $e^+e^-$\,$\to$\,$J/\psi\eta$.}
\label{Figure7}
\end{figure}
\begin{figure}\hspace{-7mm}\centerline{\epsfysize=2.6in
\epsfbox{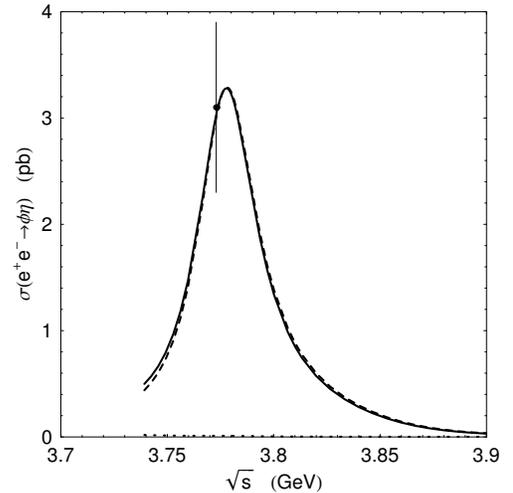}} 
\caption{The cross section for $e^+e^-$\,$\to$\,$\phi\eta$.}
\label{Figure8}
\end{figure}
The cross sections of the reactions $e^+e^-$\,$\to$\,$\gamma\chi_{
c0}$, $J/\psi\eta$, $\phi\eta$ were measured by the CLEO
Collaboration \cite{Br,Am,As} at a single point in energy
$\sqrt{s}=3773$ MeV (at the supposed maximum). Their approximate
values are presented in Table I and Figs.
\ref{Figure6}--\ref{Figure8} by the points with the error bars. They
allow us to roughly estimate the coupling constants
$g_{\psi''\gamma\chi_{c0}}\approx0.54$ GeV$^{-1}$,
$g_{\psi''J/\psi\eta}\approx0.053 $ GeV$^{-1}$, and
$g_{\psi''\phi\eta}\approx1.12\times10^{-2}$ GeV$^{-1}$, by using
Eqs. (\ref{SigmAB})--(\ref{RAB}), to construct the corresponding
cross sections. Here, as an illustration, we put $g_{\psi(2S)ab}$
and $g_{\psi''ab}$ $>0$ and $g_{\psi(2S)\gamma}/g_{\psi''\gamma}
<0$.

The solid curves in Figs. \ref{Figure6}--\ref{Figure8} show the
$e^+e^-$\,$\to$\,$\gamma\chi_{c0}$, $J/\psi\eta$, $\phi\eta$
reaction cross sections; the dashed and dotted curves show the
contributions from the $\psi''$ and $\psi(2S)$ resonances
proportional to
$$[g_{\psi''\gamma}D_{\psi(2S)}(s)+g_{\psi(2S)\gamma}\Pi_{\psi''
\psi(2S)}(s)]g_{\psi''ab}$$  and
$$[g_{\psi(2S)\gamma}D_{\psi''}(s)+g_{\psi''\gamma}\Pi_{\psi''
\psi(2S)}(s)]g_{\psi(2S)ab}$$ in Eq. (\ref{RAB}), respectively. Note
that the cross section for $e^+e^-$\,$\to$\,$\phi\eta$ is completely
dominated by the $\psi''$ contribution.

These examples tell us that the mass spectra in the $\psi''$ region
in the non-$D\bar D$ channels can be very diverse. Therefore we
should expect that the data on such spectra will impose severe
restrictions on the constructed dynamical models.


\section{CONCLUSION}

We tried to show that the shape of the $\psi''$ resonance keeps
important information about the production mechanism and
interference with background. We have considered the models
satisfying the unitarity requirement and obtained good descriptions
of the current data on the $e^+e^-\to D\bar D$ reaction cross
section, in particular, in the model with the mixed $\psi''$ and
$\psi(2S)$ resonances.

We have extracted from experiment $g^2_{\psi (2S)D\bar
D}/(4\pi)\approx13$.

Further improvement of the data and matching the results from the
different groups on the reactions $e^+e^-\to D\bar D$ can result in
crucial progress in understanding the complicate mechanism of the
$\psi''$ resonance formation.

As we have shown the measurements of the mass spectra in the
$\psi''$ region in the non-$D\bar D$ channels, such as
$e^+e^-$\,$\to$\,$\gamma \chi_{c0}$, $J/\psi\eta$, $\phi\eta$, etc.,
will also contribute to a comprehensive study of the $\psi''$
resonance physics and the effective selection of theoretical models.

Additional information about the $\psi''$ in the $D\bar D$ mass
spectra can be extracted, for example, from weak decays $B$\,$\to
$\,$\psi''K$ and photoproduction reactions at high energies $\gamma
A$\,$\to$\,$\psi''A$.

$$\mbox{\small \bf ACKNOWLEDGMENTS}$$

This work was supported in part by RFBR, Grant No. 10-02-00016, and
Interdisciplinary Project No. 102 of the Siberian division of RAS.

\begin{figure}\hspace{-7mm}\centerline{\epsfysize=2.5in
\epsfbox{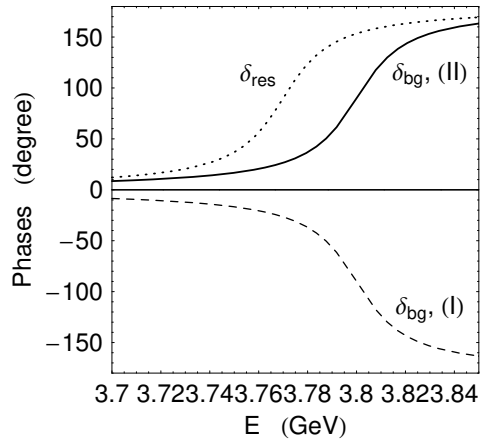}} 
\caption{The illustration of the ambiguity of the overall phase of
the $e^+e^-\to h\bar h$ reaction amplitude defined in Eq.
(\ref{A1}).} \label{Figure9}
\end{figure}

$$\mbox{\small \bf APPENDIX}$$
\setcounter{equation}{0}
\renewcommand{\theequation}{A\arabic{equation}}

If the parametrization of the reaction amplitude has no clear
dynamical justification, it can lead to unexpected problems.

Here we comment on the ambiguity of the interfering resonances
parameters determination, which has been discovered in Ref.
\cite{Bu} and discussed in Refs. \cite{To2,An2} in connection with
the $\psi''$ resonance parameters.

To illustrate the ambiguity of resonance parameters, we use a very
simple example. Consider the model $e^+e^-\to h\bar h$ reaction
amplitude (where $h$ and $\bar h$ are hadrons) involving the
resonance and background contributions \cite{Bu}:
\begin{equation}\label{A1}
F(E)=\frac{A_xe^{i\varphi_x}}{M-E-i\Gamma/2}+B_x
\end{equation}
Here $E$ is the energy in the $h\bar h$ center-of-mass system, $M$
is the mass and $\Gamma$ the energy-independent width of the
resonance, and $A_x$, $\varphi_x$, and $B_x$ are the real
parameters. At fixed $M$ and $\Gamma$, there are two solutions for
$A_x$, $\varphi_x$, and $B_x$ \cite{Bu}:
\begin{equation}\label{A2}\mbox{(I)}\quad A_x=A,\ \ B_x=B,\ \
\varphi_x=\varphi,\end{equation}
\begin{eqnarray}\label{A3}\mbox{(II)}\quad A_x=\sqrt{A^2-2AB\Gamma\sin
\varphi+B^2\Gamma^2},\ \ B_x=B,\nonumber\\
\tan\varphi_x=-\tan\varphi+B\Gamma/(A\cos
\varphi),\qquad\end{eqnarray} which yield the same cross section as
a function of energy, $\sigma(E)$\,=\,$|F(E)|^2$, and differ in the
magnitude and phase of the resonance contribution. For example, at
$M$\,=\,3.77 GeV, $\Gamma$\,=\,0.03 GeV, $A$\,=\,0.045
nb$^{1/2}$GeV, $\varphi$\,=\,0, and $B$\,=\,1.5 nb$^{1/2}$, solution
(II) gives $A_x$\,=\,$\sqrt{2}A$ and $\varphi_x$\,=\,$\pi/4$.

For each energy, the two solutions also give the different overall
phase, $\delta=\delta_{res}+\delta_{bg}$, of the amplitude $F(E)$.
For the above numerical example, the phase $\delta_{bg}$
corresponding to solutions (I) and (II) is shown in Fig.
\ref{Figure9} by the dashed and solid curves, respectively; the
phase $\delta_{res}$\,=\,$\arctan[\frac{\Gamma}{2(M-E)}]$ is shown
by the dotted curve. The origin of the rapid change of the phase
$\delta_{bg}$ (which is additional to $\delta_{res}$) requires a
special dynamical explanation (for example, the presence of extra
intermediate states), for which we do not see at present any
reasons.


\begin{thebibliography}{99}
\bibitem{FNNew} The BES Collaboration obtained also 14 values for the observed
cross section $\sigma^{\mbox{\scriptsize{obs}}}(e^+e^-\to D\bar D)$
for 3.73 GeV\,$<$\,$ \sqrt{s}$\,$<$\,3.80 GeV \cite{Ab8}. Then, the
data corrected by the initial state radiation were extracted and
analyzed in Ref. \cite{LQY}. These data are consistent within
experimental uncertainties with those used by us here.
\bibitem{FN1a} In using the $r$-dependent centrifugal barrier penetration factor
in the elementary particle physics (for example, in the case of the
$\rho$ meson \cite{PDG12}), one problem is overlooked. The existence
of $r$ implies that of potential (or background) scattering, which
is not small at the usual value of $r\approx1$ fm, both in our case
and in the case of the $\rho$ meson [recall that the phase of
$P$-wave scattering by the hard sphere of radius $r$ is
$\delta_{bg}=-rp+\arctan(rp)$ \cite{BW}, where $p$ is the scattering
momentum]. As is well known, in the $\rho$ meson region, the
background phase shift is negligible, and the $\pi\pi$ phase shift
$\delta^1_1$ is completely defined by the $\rho$ resonance.
Therefore, the descriptions of the hadronic resonance distributions
taking into account the parameter $r$ have a tentative character. As
has been recently shown in Refs. \cite{A3}, the $\pi$ meson
electromagnetic form factor can be excellently described without $r$
($r$\,=\,0) in a wide energy region
($-0.3\mbox{\,GeV}^2<s<1\mbox{\,GeV}^2$) in the field theory
approach taking into account the $\rho(770)$, $\rho(1450)$, and
$\rho(1700)$ resonances. From a theoretical point of view, the $r$
parameter does not fit into the models of Sec. III,  and we did not
introduce it. Furthermore, the data are not sensitive to its
inclusion.
\bibitem{FN2}  We omit argument $s$ of amplitudes and phases, where possible.
\bibitem{FN3}  Here we neglect the $D$ meson isovector form factor. We also do not touch
               on the fine questions about the Coulomb corrections and isospin symmetry
               breaking.
\end{thebibliography}
\end{document}